\documentclass[aps,pre,twocolumn,showpacs]{revtex4}
\usepackage{graphicx}
\begin{document}
\newcommand{\ud}{{\mathrm d}}
\newcommand{\sech}{\mathrm{sech}}

\title{Finite size fluctuations and stochastic resonance in globally coupled bistable systems}


\author{David Cubero}
\email{dcubero@us.es}
\affiliation{F\'{\i}sica Te\'orica, Universidad de Sevilla, Apartado de Correos 1065, Sevilla 41080, Spain}


\date{\today}

\begin{abstract}
The dynamics of a system formed by a finite number $N$ of globally coupled bistable oscillators and driven by external forces is studied focusing on a global variable defined as the arithmetic mean of each oscillator variable. Several models based on truncation schemes of a hierarchy of stochastic equations for a set of fluctuating cumulant variables are presented. This hierarchy is derived using It\^o stochastic calculus, and the noise terms in it are treated using an asymptotic approximation valid for large $N$. In addition, a simplified one-variable model based on an effective potential is also considered. These models are tested in the framework of the phenomenon of stochastic resonance. In turn, they are used to explain in simple terms the very large gains recently observed in these finite systems. 
\end{abstract}


\maketitle

\section{Introduction}
Noise induced phenomena in nonlinear systems have attracted a great deal of attention in a variety of contexts in physics, chemistry and the life sciences. 
 An important example is the phenomenon of stochastic resonance (SR) \cite{gamhan98}, in which the response of the system (output) to external driving (input) is amplified and optimized for certain values of the noise parameters. More specifically, the non-monotonic behavior of the output signal-to-noise ratio (SNR) with the strength of the noise is a widely accepted signature of SR. In addition, a dimensionless quantity known as the SR gain is usually defined as the ratio of the output SNR over the input SNR.

Very recently, very large SR gains have been reported for systems formed by a finite number $N$ of globally coupled bistable oscillators \cite{casgom06,cubcas07}. Here the term {\em global coupling} is used to indicate that each oscillator interacts with {\em all} other oscillators. These systems were used years ago by Kometani and Shimizu \cite{komshi75} as an empirical model to describe muscle contraction. Later on, Desai and Zwanzig \cite{deszwa78} gave a more detailed statistical mechanical description, finding an order-disorder transition for a variable defined as the expectation value of the position of one oscillator. This variable is used to study the global behavior of the coupled bistable system. Desai and Zwanzig focused on systems with infinitely large sizes, $N\rightarrow\infty$, investigating the system dynamics by analyzing the time evolution of the central cumulant moments of one-oscillator's distribution. In addition, a Gaussian approximation was proposed in order to close the cumulant moment hierarchy and obtain analytical expressions. A similar approach is currently used as a mean field approximation in the investigation of various noise-induced phenomena such as noise-induced phase transitions \cite{kawsai04,sagsan07}. Recently, in order to study the effect of fluctuations due to the finite size of the system, Pikovsky et al.~\cite{pikzai02} extended this approach by replacing the expectation values of one-oscillator properties $\langle\cdot\rangle$ by arithmetic means over all oscillators $N^{-1}\sum_{i=1}^N(\cdot)$. A Gaussian approximation, including noisy terms, was derived and used to illustrate the phenomenon of system size resonance, in which the SR quantifiers display a non-monotonic behavior as a function of $N$. In this paper, the work by Pikovsky et al.~is extended to higher orders in the fluctuating cumulant dynamics. The Gaussian approximation is re-derived using a rigorous formalism based on It\^o stochastic calculus and compared with other approximations.  

One important goal of this paper is to explain the very large gain values observed in globally coupled bistable systems \cite{casgom06,cubcas07}, especially when compared with those observed in uncoupled or isolated bistable systems. To that effect, it is desirable to derive a simplified theory in which the number of degrees of freedom is much smaller than the number of coupled oscillators, thus being more amenable to analytical treatment or qualitative interpretation. In this regard, the Gaussian approximation is a practical alternative, though in principle not fully satisfactory, because it is not based on a small parameter expansion but on an uncontrolled assumption (the neglect of cumulants higher than the second) that is known to be not accurate even in the limit of an infinite system \cite{deszwa78} . 

In this paper, this approximation is presented, as well as other simplified models with a reduced number of degrees of freedom which are able to mimic the most important features of the system dynamics with a finite size. These simplified models represent different approximation schemes and might be regarded as an expansion or generalization of the work by Desai and Zwanzig \cite{deszwa78} and Pikovsky et al.~\cite{pikzai02}.

The paper is organized as follows. In the next section, the model system  and the SR quantifiers are defined. The simplified models are presented in Sec.~\ref{sec:eff}. These models are compared to the original model system by means of computer simulations in Sec.~\ref{sec:sim}. Finally, Sec.~\ref{sec:con} provides a short summary and conclusions.

\section{Model and definitions}
Let us consider a set of $N$ interacting bistable oscillators, each one of them characterized by a single degree of freedom $x_i$ ($i=1,\ldots,N$), whose dynamics is governed by the Langevin equations \cite{komshi75,deszwa78}
\begin{equation}
\dot{x}_i=x_i-x_i^3+\frac{\theta}{N}\sum_{j=1}^N(x_j-x_i)+\xi_i(t)+F(t),
\label{eq:lang}
\end{equation}
where $\xi_i(t)$ is a Gaussian white noise with zero average and 
\begin{equation}
\langle \xi_i(t)\xi_j(s)\rangle=2D\delta_{ij}\delta(t-s),
\label{eq:xicorr}
\end{equation}
 $\theta$ is a coupling parameter defining the strength of the interaction between oscillators, and $F(t)$ is an external driving force of period $T$. 

To characterize the system as a whole we define the {\em collective} or {\em global} variable $S(t)$ as
\begin{equation}
S(t)=\frac{1}{N}\sum_{i=1}^N x_i(t).
\label{eq:def:S}
\end{equation}
The stochastic resonance quantifiers for this variable are defined in the usual way. The one-time correlation function,
\begin{equation}
C(\tau)=\frac{1}{T}\int_0^T \ud t\langle S(t+\tau)S(t)\rangle_{\infty},
\end{equation}
can be written as the sum of two contributions: a coherent part, 
\begin{equation}
C_\mathrm{coh}(\tau)=\frac{1}{T}\int_0^T \ud t\langle S(t+\tau)\rangle_{\infty}\langle S(t)\rangle_{\infty},
\end{equation}
which is periodic with period $T$, and an incoherent part 
\begin{equation}
C_{\mathrm{incoh}}(\tau)=C(\tau)-C_{\mathrm{coh}}(\tau),
\end{equation}
which decays to zero for large values of $\tau$ and reflects the
correlation of the process $S(t)$ at different times due to
fluctuations. In the expressions above, the notation $\langle \cdots \rangle $ indicates an average over the noise realizations and 
the subscript ``$\infty$'' indicates the long time limit of the noise average, i.~e., its value after waiting for $t$ long enough that transients have died out.
The SNR of a random signal measures the signal strength relative to its background noise. More specifically, we calculate the output SNR as 
\begin{equation}
R_\mathrm{out}=\frac{Q_u}{Q_l},
\end{equation}
where 
\begin{equation}
Q_u=\frac{2}{T}\int_0^T \ud\tau\, C_{\mathrm{coh}}(\tau)\cos(\Omega\tau),
\end{equation}
$\Omega=2\pi/T$ being the driving frequency, and
\begin{equation}
Q_l=\frac{2}{\pi}\int_0^\infty \ud\tau\, C_{\mathrm{incoh}}(\tau)\cos(\Omega\tau).\end{equation}
Note that the quantity $Q_{u}$ is proportional to the so-called spectral amplification, which is another widely used SR quantifier.

As the size of the system $N$ is increased while keeping the interaction parameter $\theta$ constant, the collective variable $S(t)$ becomes less noisy, becoming completely deterministic in the limit $N\rightarrow\infty$. As a result, $R_\mathrm{out}$ diverges in that limit. This is a consequence of the averaging process implicit in the definition (\ref{eq:def:S}).  

The SR gain is defined as 
\begin{equation}
G=\frac{R_\mathrm{out}}{R_\mathrm{in}},
\label{eq:gaindef}
\end{equation}
where $R_\mathrm{in}$ is the SNR of the collective input signal $N^{-1}\sum_{i=1}^N[ F(t)+\xi_i(t)]$. For example, for a periodic rectangular driving force of amplitude $A$, the input SNR is given by $R_\mathrm{in}=4A^2N/(\pi D)$. The SR gain (\ref{eq:gaindef}) is a dimensionless quantity that measures the amplification of the system response with respect to the collective input signal. The input SNR $R_\mathrm{in}$ diverges linearly with $N$ in the limit $N\rightarrow\infty$, so that the SR gain remains finite. 

Since in a system with coupled linear oscillators the SNR of the collective process equals the SNR of the collective input signal, i.e. $R_\mathrm{out}^{(L)}=R_\mathrm{in}$, the SR gain also measures the response of the non-linear system with respect to that of a linear system subject to the same deterministic and stochastic forces.

Additionally, in the absence of interaction between the bistable oscillators (the case $\theta=0$), the collective SR gain $G$ equals the SR gain of each  independent oscillator \cite{casgom06}. Thus, by comparing the SR gain values of the collective variable of a finite set of interacting oscillators with those observed in the case $N=1$, we have a useful tool to highlight nonlinear effects that are a direct consequence of the coupling between the oscillators. 

\section{Finite size dynamics}
\label{sec:eff}
In this section, we define a set of stochastic processes, which we will refer to as {\em fluctuating cumulants}, in order to describe the dynamics of a finite system of coupled oscillators in terms of a reduced number of variables. 
Then, by using It\^o stochastic calculus, we derive the hierarchy of equations that these cumulants obey. A few approximation schemes are proposed for systems with a large but finite number of oscillators. Finally, we introduce a simple one-variable model in which the dynamics of $S(t)$ is mimicked by using an effective potential. 

The fact that the infinite system ($N=\infty$) is completely deterministic, and the approximations described in this section are valid for large $N$, makes these methods especially appropriate to study the effect of fluctuations due to the finite size of the system. 

\subsection{Fluctuating cumulants}
Let us define the set of stochastic variables
\begin{equation}
M_k(t)=\frac{1}{N}\sum_{i=1}^N[x_i(t)-S(t)]^k,
\end{equation}
with $k$ being a positive integer.
We will refer to $M_k$ as the {\em fluctuating moment} of order $k$. Note that  $M_1(t)=0$. 

In order to obtain a hierarchy of stochastic differential equations for these variables, we need first to choose a convenient stochastic interpretation. The Langevin equations (\ref{eq:lang}) are well defined and do not depend on the stochastic interpretation. Note, however, that the specific form of the stochastic differential equations for the fluctuating moments do depend on the stochastic calculus utilized. In the following, unless explicitly stated, It\^o stochastic calculus is assumed. It is customary within this calculus to use a notation to express the stochastic differential equations in which there is no explicit mention to the white noises (see for example \cite{oksendal03}). In particular, Eq.~(\ref{eq:lang}) would be written as
\begin{equation}
\ud{x}_i=[x_i-x_i^3+\frac{\theta}{N}\sum_{j=1}^N(x_j-x_i)+F(t)]dt+(2D)^{1/2}\ud B_i,
\label{eq:lang:ito}
\end{equation}
where $\ud B_i$, with $i=1,\ldots, N$, is the differential of the Wiener process $B_i(t)$ with properties 
\begin{equation}
\ud B_i(t)\ud B_j(t)=\delta_{ij}dt.
\label{eq:wienner}
\end{equation}
The Gaussian white noise $\xi_i(t)$ can be viewed as proportional to the derivative of $B_i(t)$, $\xi_i=(2D)^{1/2}\ud B_i/\ud t$, though it is not an ordinary  stochastic process but a generalized process and requires a special formalism to be defined rigorously (see \cite{oksendal03} and references within). Here we will use both notations at convenience. 

 Using It\^o differentiation rules \cite{oksendal03} we find the following stochastic differential equations for the fluctuating moments
\begin{eqnarray}
\frac{\dot{M}_k}{k} &=& (1-3S^2-\theta)M_k-M_{k+2}-3SM_{k+1} \nonumber\\
& &+(M_3+3SM_2)M_{k-1}+(k-1)D\left(1-\frac{1}{N}\right) \nonumber\\
& &-\eta M_{k-1}+\mu_{k-1},
\label{eq:hier:mk}
\end{eqnarray}
where 
\begin{equation}
\mu_k(t)=\frac{1}{N}\sum_{i=1}^N[x_i(t)-S(t)]^k\xi_i(t)
\label{eq:muk:def}
\end{equation}
and
\begin{equation}
\eta(t)=\mu_0(t)=\frac{1}{N}\sum_{i=1}^N\xi_i(t)
\end{equation}
are $\delta-$correlated noises with the following first moments 
\begin{eqnarray}
\langle \eta(t)\rangle &=& \langle \mu_k(t)\rangle = 0, \label{eq:muk:0}\\
\langle \eta(t) \eta(t^\prime)\rangle&=&\frac{2D}{N}\delta(t^\prime-t), \label{eq:etaeta}\\
\langle \mu_k(t) \mu_{k^\prime}(t^\prime)\rangle&=&\frac{2D}{N}\delta(t^\prime-t)\langle M_{k+k^\prime}(t)\rangle.
\label{eq:mukmukp}
\end{eqnarray}
Notice that the result (\ref{eq:mukmukp}) is only obtained from (\ref{eq:muk:def}) when  It\^o calculus is assumed (see Appendix \ref{sec:apen1}). Using Stratonovich calculus leads to a much more intricate expression in which the approach proposed in this paper is not applicable.  

Additionally, note that (\ref{eq:mukmukp}) implies that the processes $\mu_k$ are not uncorrelated. Rigorously, only $\eta(t)$ is a Gaussian process. Nevertheless, it can be shown that in the asymptotic limit of a very large number of oscillators, $N\rightarrow\infty$, all $\mu_k$ tend towards a Gaussian behavior (see Appendix \ref{sec:apen2}). This property will allow us to rewrite Eq.~(\ref{eq:hier:mk}) as a closed set of stochastic equations for the fluctuating variables in that limit.

We can define a set of fluctuating cumulants by using the formula:
\begin{equation}
K_n=M_n-\sum_{k=1}^{n-1}\frac{(n-1)!}{k!(n-1-k)!}K_{n-k}M_{k}.
\label{eq:mom:cum}
\end{equation}
Equation (\ref{eq:mom:cum}) is the formula that relates the moments with the cumulant moments of a single-variable stochastic distribution. When the stochastic variable is Gaussian, all cumulants $K_n$, with $n\ge3$, exactly vanish. A description in terms of cumulants is preferable because, unlike a descriptions with moments, it is expected that higher order cumulants are negligible in comparison with lower order cumulants, especially if the deviation with respect a Gaussian behavior is not very large. 
 
In terms of the fluctuating cumulants $K_n$, the first three equations of the hierarchy (\ref{eq:hier:mk}) are 
\begin{eqnarray}
\dot{S}&=&S-S^3-3S K_2-K_3+\eta+F(t), \label{eq:hie:cum:1}\\
\frac{\dot{K_2}}{2}&=&(1-3S^2-\theta) K_2-3S K_3-3K_2^2-K_4 \nonumber \\
& &+D(1-\frac{1}{N})+\mu_1, \label{eq:hie:cum:2}\\
\frac{\dot{K_3}}{3}&=&(1-3S^2-\theta) K_3-3S(2 K_2^2+K_4)\nonumber \\
& &-9 K_2K_3-K_5 -K_2\eta+\mu_2. 
\label{eq:hie:cum:3}
\end{eqnarray}

The noise terms $\eta$ and $\mu_k$ vanish in the formal limit $N\rightarrow\infty$, therefore leading to a deterministic hierarchy of equations for the fluctuating moments or cumulants. This deterministic hierarchy is equivalent to the non-linear hierarchy obtained by Desai and Zwanzig in \cite{deszwa78} for the cumulant moments of the process $y_1(t)=x_1(t)-S(t)$. In contrast to the theory presented in Ref.~\cite{deszwa78}, which is based on the calculation of one-time expectation values, the fluctuating cumulant approach will allow us to study dynamical properties such as autocorrelation functions. 

When $N$ is finite, the hierarchy of equations that $M_k$ or $K_k$ obeys is not closed, since the noise processes $\mu_k$ depend on $M_k$ in a non-trivial way. In the next subsections we present a few approximative schemes that overcome this difficulty for systems with a large number of coupled oscillators.  

\subsection{Second order approximation}
If we retain the first two equations of (\ref{eq:hie:cum:1})-(\ref{eq:hie:cum:3}), neglect all fluctuating cumulants $K_n$ with $n\ge 3$, and also neglect the term $1/N$ and the noise $\mu_1(t)$ in (\ref{eq:hie:cum:2}), we obtain a closed set of equations for the processes $S(t)$ and $K_2(t)$:
\begin{eqnarray}
\dot{S}&=&S-S^3-3S K_2+\eta+F(t), \nonumber\\
\frac{\dot{K_2}}{2}&=&(1-3S^2-\theta) K_2-3K_2^2+D. 
\label{eq:pikovsky}
\end{eqnarray}
These set of equations was proposed by Pikovsky et al.~in Ref.~\cite{pikzai02}. This truncation scheme has been called ``the Gaussian approximation'' because all fluctuating cumulants with order higher than the second one are neglected. There is no reason to expect a priori that these higher order cumulants can be neglected in any limit, other than the hope that their contribution is small. Note, in addition, that in this scheme the $\delta$--correlated noise $\mu_1(t)$ is neglected without justification.  

\subsection{Third order approximation}
Let us now focus in a third order truncation scheme. We will retain the three Eqs. (\ref{eq:hie:cum:1})-(\ref{eq:hie:cum:3}), but consistently neglect $K_4$ and $K_5$. 

Since each $\mu_k(t)$ for $k=1,2,\ldots$ is a Gaussian process in the lowest order in $N^{-1}$ (see Appendix \ref{sec:apen2}), its probability distribution is completely determined by its first moments (\ref{eq:muk:0})--(\ref{eq:mukmukp}). As mentioned before, the processes $\mu_k(t)$ are not independent of each other. Thus, it is preferable to express them in terms of a set of independent Gaussian noises $\eta_l(t)$ with zero mean and  
\begin{equation}
\langle \eta_l(t)\eta_{l^\prime}(t^\prime)\rangle=\frac{2D}{N}\delta_{l l^\prime}\delta(t-t^\prime),
\end{equation}
where $l,l^\prime\ge 0$ and $\eta_0\equiv\eta$.
With the expansion
\begin{equation}
\mu_k=\sum_{l=0}^k c_{kl}\,\eta_l,
\end{equation}
we only need to select the coefficients $c_{kl}$ so that the correlations (\ref{eq:mukmukp}) are satisfied. This can be achieved by using the Gram-Schmidt ortho-normalization method. The result for the first two terms is
\begin{eqnarray}
\mu_1&=&\langle K_2\rangle^{\frac{1}{2}}\eta_1 \label{eq:mu1}\\
\mu_2&=&\langle K_2\rangle\eta+\frac{\langle K_3\rangle}{\langle K_2\rangle^{\frac{1}{2}}}\eta_1+ \nonumber \\
& {} & +\frac{|\langle K_4\rangle \langle K_2\rangle+2\langle K_2\rangle^3-\langle K_3\rangle^2|^{\frac{1}{2}}}{\langle K_2\rangle^{\frac{1}{2}}}\eta_2.
\label{eq:mu2}
\end{eqnarray}
Note, however, that in these expressions the coefficients $c_{kl}$ appear as functions of the average values of the fluctuating cumulants $K_n$. Thus, if we plan to solve the Eqs. (\ref{eq:hie:cum:1})--(\ref{eq:hie:cum:3}) using (\ref{eq:mu1})--(\ref{eq:mu2}), we would have to consider the equation of motion for $\langle K_n\rangle$ \cite{deszwa78} and solve the whole set of equations self-consistently. Alternatively, we could use a slightly different version of (\ref{eq:mu1})--(\ref{eq:mu2}) in which the expected values $\langle K_n\rangle$ are replaced by $K_n$. This way, the correlations (\ref{eq:mukmukp}) for the first noise terms are also identically satisfied, and the fact that the fluctuating cumulants become deterministic in the limit $N\rightarrow\infty$ guarantees that the proposed expressions for the noises $\mu_k(t)$ are Gaussian in the lowest order in $N^{-1}$. 
Since all Gaussian processes are completely determined by its first two moments, both methods to generate the noise sources $\mu_k$ are mathematically equivalent in the asymptotic limit of large $N$, though the later is more physically appealing because in this case the instantaneous value of the noise source $\mu_k(t)$ in one trajectory does not depend on averages over trajectories but on single-trajectory values. 

Using this later procedure, the following stochastic differential equations with multiplicative noise are obtained 
\begin{eqnarray}
\dot{S}&=&S-S^3-3S K_2-K_3+\eta+F(t), \nonumber\\
\frac{\dot{K_2}}{2}&=&(1-3S^2-\theta) K_2-3S K_3-3K_2^2 \nonumber \\
& &+D+|K_2|^{\frac{1}{2}}\eta_1, \nonumber\\
\frac{\dot{K_3}}{3}&=&(1-3S^2-\theta) K_3-6S K_2^2-9 K_2K_3 \nonumber \\
& &+\frac{K_3\eta_1+ |2K_2^3-K_3^2|^{\frac{1}{2}}\eta_2}{|K_2|^{\frac{1}{2}}}.
\label{eq:third}
\end{eqnarray}
This method has also the advantage that the system of equations (\ref{eq:third}) can be solved numerically using standard stochastic algorithms \cite{kloepla92}. It represents a third order approximation scheme. Notice that this scheme can be applied in a straightforward way to obtain the corresponding equations of an arbitrary order truncation of the fluctuating cumulant hierarchy.  

\subsection{Effective potential}
As we increase the truncation order of the fluctuating cumulant hierarchy, as we have discussed above, a more accurate approximation is obtained. However, the number of equations is also increased. On the other hand, we may wonder how good a description based on a single differential equation is. The aim is to derive a simplified model that may not mimic quantitatively but qualitatively the coupled system dynamics, in addition to being more amenable to analytical treatment.   

Here we consider the following single stochastic equation
\begin{equation}
\dot{S}=-U_{\mathrm{eff}}'(S)+\eta+F(t),
\label{eq:eff}
\end{equation}
where $\eta(t)$ is the Gaussian white noise defined by (\ref{eq:muk:0})--(\ref{eq:etaeta}), and $U_{\mathrm{eff}}(S)$ an effective potential to be specified.

\begin{figure}
\includegraphics[width=7.cm]{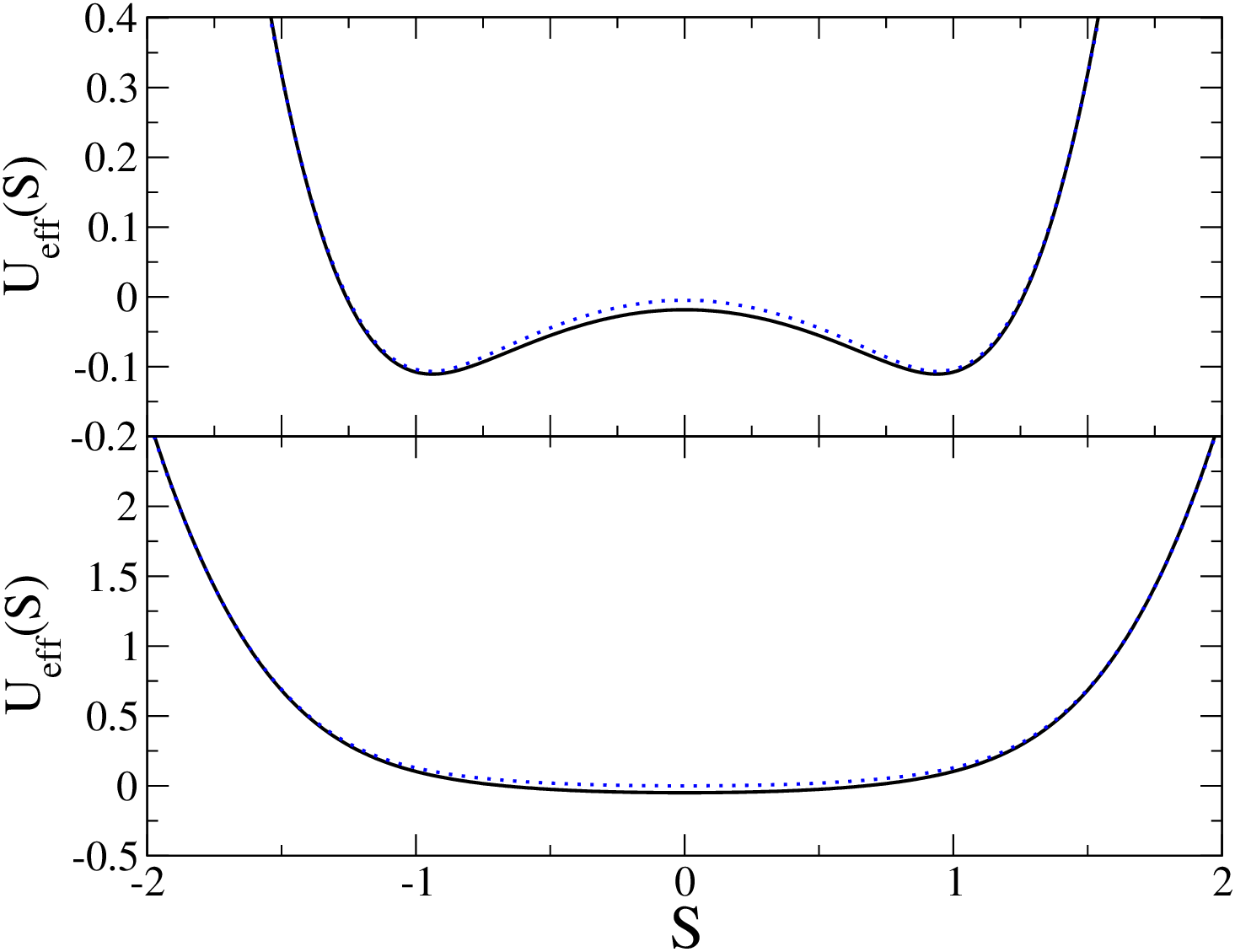}
\caption{
\label{fig:Ueff} 
Effective potential for the simple Langevin model (\ref{eq:eff}) for two systems with $\theta=0.5$. Top panel corresponds to $D=0.08$ ($<D_c\approx0.2645$) and the bottom panel to $D=0.4$ ($>D_c$). The solid lines depict the analytical solution given in Ref.~\cite{deszwa78}, whereas the dotted lines correspond to the effective potential obtained using the simulation method described in the text for a system with $N=10$.
}
\end{figure}

We can determine the effective potential uniquely by requiring the model to reproduce the equilibrium properties of the original system. The stationary probability density $P_\mathrm{eq}(S)$ of the Langevin Eq.~(\ref{eq:eff}) in the absence of external driving is given by \cite{vankampen}
\begin{equation}
P_\mathrm{eq}(S)=Z^{-1}\exp\left(-\frac{N U_\mathrm{eff}(S)}{D}\right),
\label{eq:peq}
\end{equation}
where $Z$ is a normalization constant. Therefore, by inverting (\ref{eq:peq}), we find an expression for the effective potential up to an additive constant $c$,
\begin{equation}
U_\mathrm{eff}(S)=-\frac{N}{D} \ln P_\mathrm{eq}(S)+c.
\end{equation}
In Ref.~\cite{deszwa78}, Desai and Zwanzig presented an analytical expression for the equilibrium density $P_\mathrm{eq}(S)$ by retaining the leading term in the asymptotic expansion for large $N$. We will refer as $U_\mathrm{eff}^{(\infty)}(S)$ to the corresponding effective potential. This analytical solution shows that for a given value of $\theta$ there exists a $D=D_c$ such that for values $D$ greater than $D_c$, the effective potential $U_\mathrm{eff}^{(\infty)}(S)$ is monostable with a minimum located at $S_0=0$. For $0\le D<D_c$, the effective potential is bistable with two minima at $\pm S_0$, being $S_0$ a function of $\theta$ and $D$. Figure~\ref{fig:Ueff} depicts this situation for a system with $\theta=0.5$. The calculated critical noise for this value of $\theta$ is $D_c\approx 0.2645$ \cite{deszwa78}.

For a system with a finite size $N$, we can calculate numerically $P_\mathrm{eq}(S)$ by simulating the Langevin equations (\ref{eq:lang}) and computing the histogram of the collective variable $S(t)$ after a sufficiently large time when the system has equilibrated. Figure \ref{fig:Ueff} shows the resulting effective potential $U_\mathrm{eff}^{(N)}(S)$ for a system with $N=10$ oscillators. It can be seen that the deviations with respect to the infinite size potential $U_\mathrm{eff}^{(\infty)}(S)$ are very small, even for such a small system.

\section{Stochastic resonance revisited}
\label{sec:sim}

\begin{figure}
\includegraphics[width=8.5cm]{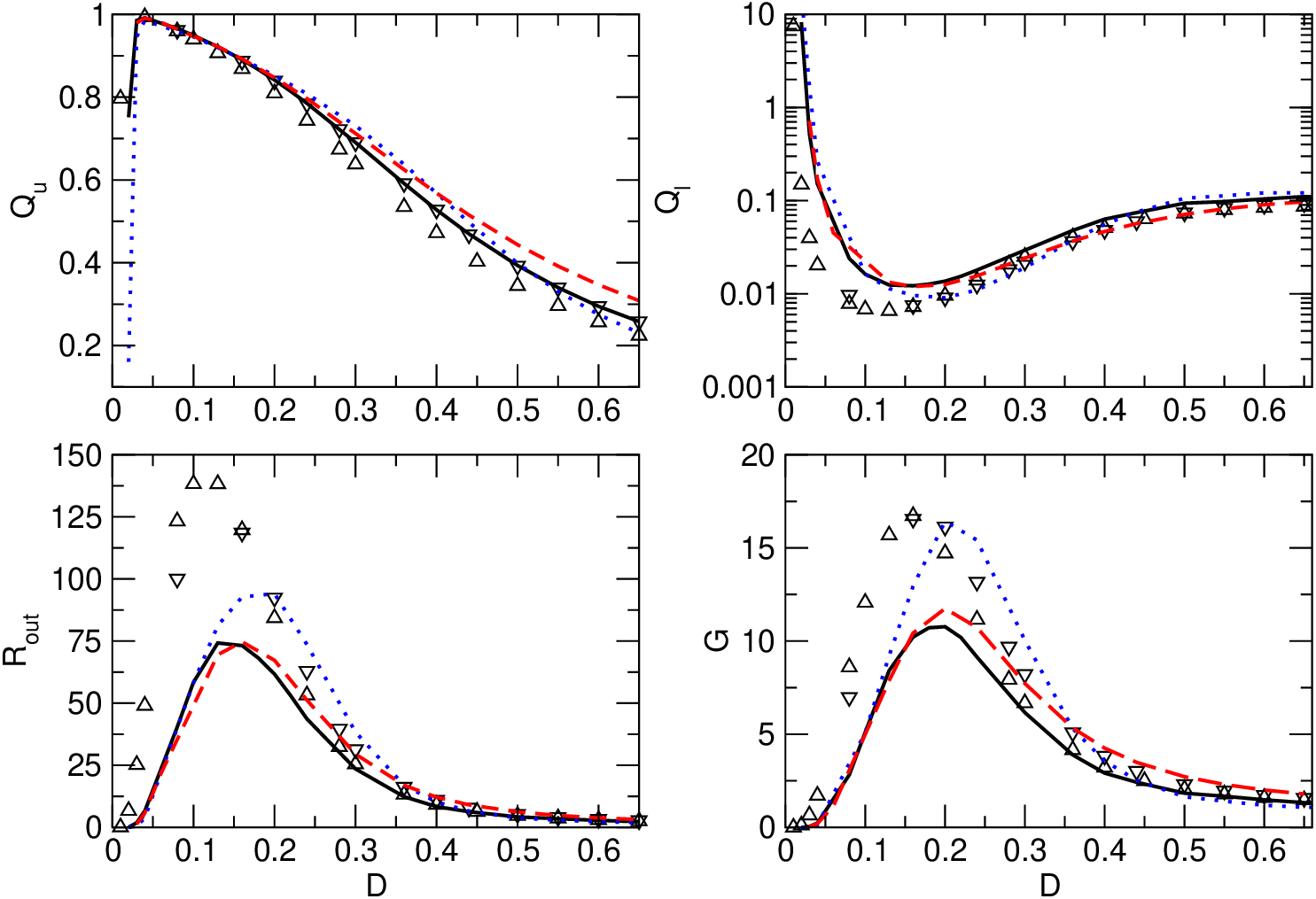}
\caption{
\label{fig:N10} 
Effective dynamics for a coupled system with $N=10$ oscillators subject to a rectangular driving of amplitude $A=0.3$ and frequency $\Omega=0.01$. Several stochastic resonance quantifiers are depicted vs the noise strength $D$. From top-left to right-bottom: the numerator of the SNR ($Q_u$), the denominator ($Q_l$), the output SNR ($R_\mathrm{out}$) and the gain ($G$). The solid lines depict the numerical solution of the full Langevin dynamics (\ref{eq:lang}). The dotted and dashed lines correspond to the fluctuating cumulants approach truncated at the second (\ref{eq:pikovsky}) and third order (\ref{eq:third}), respectively. The results for the effective potential approach (\ref{eq:eff}) are depicted by triangles pointing upwards ($U_\mathrm{eff}^{(\infty)}$) and downwards ($U_\mathrm{eff}^{(10)}$). 
 }
\end{figure}

In this section, we compare numerically the predictions of the effective models presented in Sec.~\ref{sec:eff} in the framework of SR. The simplified character of these models will allow us to explain in intuitive terms the highly nonlinear effects observed in the stochastic resonance quantifiers \cite{casgom06,cubcas07}. 

We will restrict our study to a periodic rectangular driving force,  
\begin{equation}
F(t)=(-1)^{n(t)} A,
\label{eq:driving}
\end{equation}
where $n(t)=\lfloor 2t/T\rfloor$, $\lfloor z\rfloor$ being the floor function of $z$. The input SNR for forces of this type can be readily calculated as $R_\mathrm{in}=4A^2N/(\pi D)$.
In all cases reported here the coupling strength is fixed to $\theta=0.5$, the driving frequency $\Omega=2\pi/T$ to $\Omega=0.01$, and the driving amplitude to $A=0.3$. This amplitude is {\em subthreshold} in the sense that the driving force (\ref{eq:driving}) cannot induce sustained oscillations between the different attractors of the dynamics in the absence of noise (i.e., for $D=0$).

The stochastic differential equations presented in the preceding sections were solved numerically by using weak predictor-corrector algorithms of order 2.0 \cite{kloepla92}. 

Figure~\ref{fig:N10} shows several SR quantifiers as a function of the noise strength $D$ for a coupled system with $N=10$ oscillators. A strong amplification of the collective response is observed, with SR gains reaching very large values, especially when compared with uncoupled systems subject to the same input signals (see Ref.~\cite{casgom03}). These findings were first reported in \cite{casgom06}. Since the numerator $Q_{u}$ of the SNR remains of the same order of magnitude for the range of noise strength values plotted, the large values of the SR gain are mainly due to the reduction of a few orders of magnitude of the denominator $Q_l$, as shown in the top-right panel of Fig.~\ref{fig:N10}. 

The SR quantifiers obtained with the effective potential model described by Eq.~(\ref{eq:eff}) are depicted by triangles in Fig.~\ref{fig:N10}. Triangles pointing upwards correspond to the effective potential $U_\mathrm{eff}^{(\infty)}$ in the asymptotic limit $N\rightarrow\infty$, whereas triangles pointing downwards correspond to the effective potential $U_\mathrm{eff}^{(10)}$ computed numerically for a system with $N=10$. In Fig.~\ref{fig:N10}, it can be seen that the later leads to a better agreement than the former for $Q_u$ due to the small but appreciable discrepancies observed in Fig.~\ref{fig:Ueff}. However, no significant improvement is seen in the rest of the quantifiers: $Q_l$, the SNR, and the gain. In general, the effective potential approach is able to describe qualitatively the phenomenon, displaying a non-monotonic behavior with a maximum at about the same value of the noise strength $D$ than the original system. Nevertheless, quantitatively the agreement is not so good, showing a consistent underestimation of the noise term $Q_l$ by roughly a factor of 2. The fact that this approach underestimates the fluctuations of the collective variable is easy to understand if one takes into account that the effective potential is a mean-field-like idealization in which the real discrete, more noisy, interaction is replaced by a smoothed potential.

A slightly better quantitative agreement is obtained with the Gaussian approximation described by Eq.~(\ref{eq:pikovsky}), i.e. the fluctuating cumulants approach truncated at the second order,  which is depicted in Fig.~\ref{fig:N10} by dotted lines. It can be seen that the SNR is in better agreement, though the SR gain around the maximum has not been improved significantly overall. A considerably enhanced agreement is achieved by the third order approach in Eq.~(\ref{eq:third}), which is represented in Fig.~\ref{fig:N10} by dashed lines. It can be seen that this third order approximation slightly underestimates the noise term $Q_l$ for large enough values of $D$. This is what one would expect, because this method neglects the higher order cumulants $K_4$ and $K_5$ in Eq.~(\ref{eq:third}) [see Eqs.~(\ref{eq:hie:cum:2})--(\ref{eq:hie:cum:3})], which, if present, would increase the fluctuations of the lower order cumulants. 
 
\begin{figure}
\includegraphics[width=8.5cm]{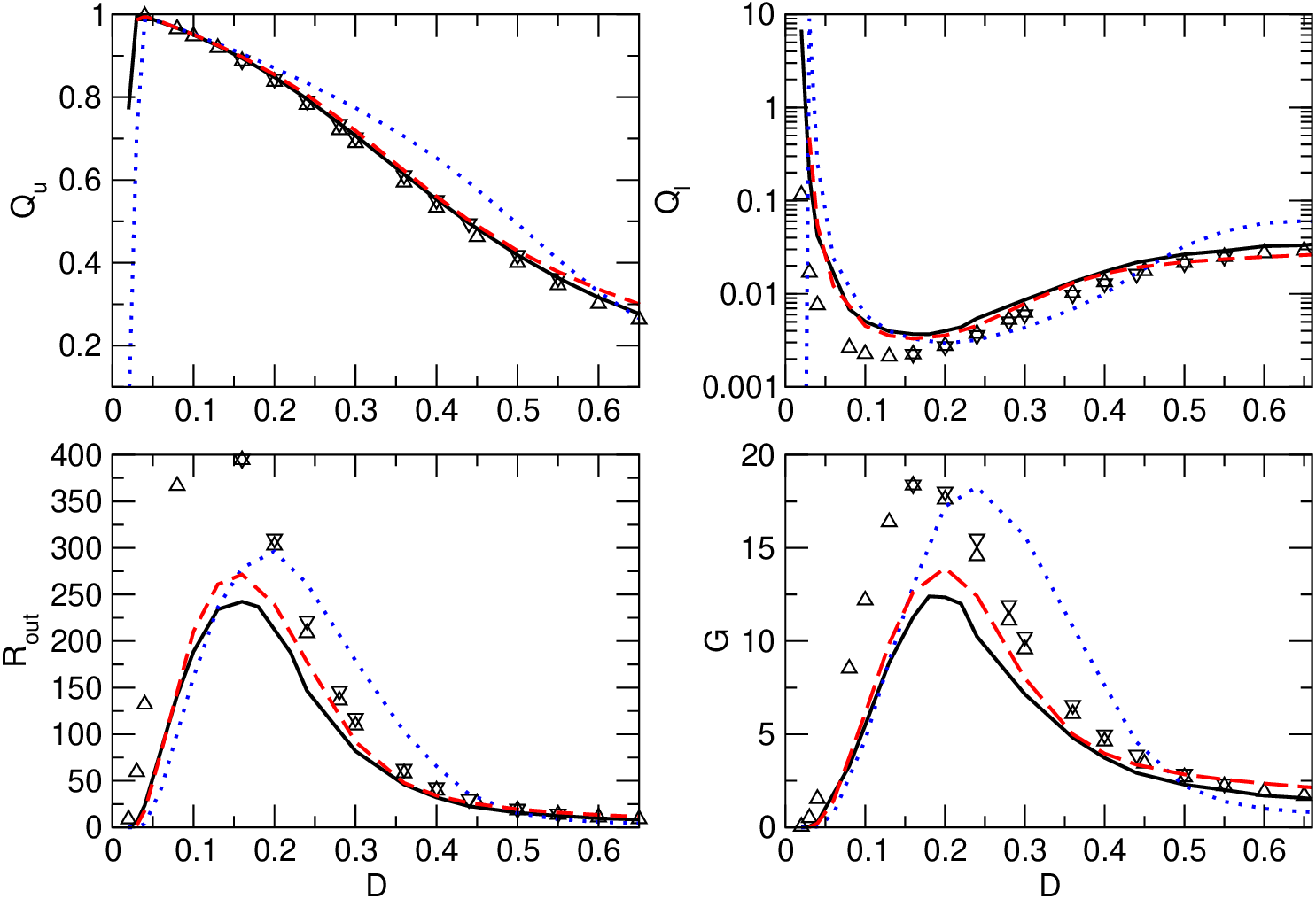}
\caption{
\label{fig:N30} 
Same as in Fig.~\ref{fig:N10} but for a system with $N=30$. 
}
\end{figure}

Figure \ref{fig:N30} confirms the above discussed behavior of the simplified models for a larger system with $N=30$. Again, the effective potential theory and the Gaussian approximation provide mainly a qualitative picture, with quantitative predictions within the same order of magnitude. The best quantitative agreement is also observed to be given by the third-order fluctuating cumulant scheme.  The main differences with respect to the smaller system discussed before are in the quantity $Q_u$, which is proportional to the spectral amplification. In this case the system is large enough so that very small differences are observed between the effective potential $U_\mathrm{eff}^{(30)}$ and  $U_\mathrm{eff}^{(\infty)}$. They both provide data in very good agreement with the original system data. In addition, notice that the Gaussian approximation data for $Q_u$ deviates appreciably from the system data for large enough values of $D$. The fact that the third order approximation leads to a good agreement for $Q_u$ indicates that the third cumulant plays an important role for the spectral amplification at these noise strength values.

Finally, we now use these simplified models to explain the very large gain values observed in globally coupled bistable systems \cite{casgom06,cubcas07}, and particularly those observed in the bottom-right panels of Figs.~\ref{fig:N10} and \ref{fig:N30}. These gain values are much larger than those observed in uncoupled or isolated bistable systems subject to the same rectangular input signals \cite{casgom03}, and thus are due to the interaction between the oscillators. For the sake of simplicity we will use the effective potential approximation in the following discussion, though similar arguments can be used within the other schemes presented in Sec.~\ref{sec:eff}. 
The only noisy term in Eq.~(\ref{eq:eff}) is $\eta(t)$, which has a strength of $D/N$. Thus, we can fix the individual noise strength $D$ and still being able to get rid of the noise by considering the limit $N\rightarrow\infty$. This way Eq.~(\ref{eq:eff}) becomes deterministic and we can apply the concept of static threshold for a finite value of $D$. A simple analysis of the infinite size potential shows that a constant driving of $A=0.3$ is able to remove one of the two attractors of the dynamics for systems with noise strength values $D$ larger than $D_c(A\!=\!0.3)\approx 0.02$ (and, of course,  smaller than $D_c(A\!=\!0)=D_c\approx0.2645$, the noise value where the effective potential turns monostable in the absence of driving). Therefore, most of the data points in Figs.~\ref{fig:N10} and \ref{fig:N30} correspond to {\em suprathreshold} dynamics when viewed from the perspective of the Langevin equation (\ref{eq:eff}). The response is expected to be more amplified with suprathreshold signals than with subthreshold signals, because, for the first ones, the presence of noise is not necessary in order to produce jumps between the two locations of the time-dependent attractor. For instance, only gains larger than unity has been found for an isolated bistable system subject to monochromatic signals when the driving amplitude is suprathreshold \cite{haninc00}, with gain values reaching a few tenths above unity. In fact, the above consideration may well explain why gains larger than unity (also a few tenths above unity) are found for globally coupled bistable systems subject to monochromatic signals \cite{casgom06}: the collective variable dynamics is effectively suprathreshold.

\begin{figure}
\includegraphics[width=8.5cm]{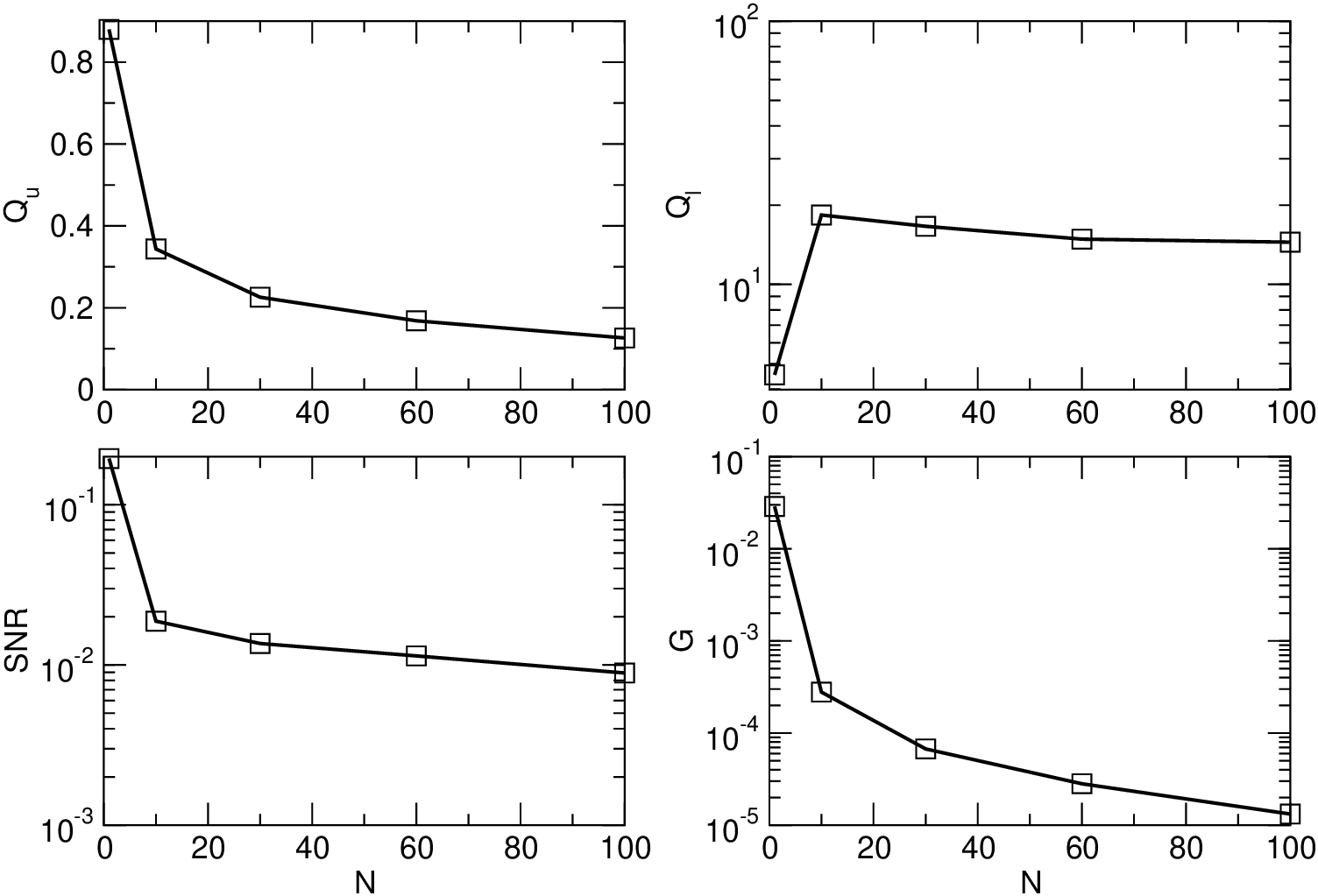}
\caption{
\label{fig:sub}
Subthreshold dynamics. Same stochastic resonance quantifiers as in Fig.~\ref{fig:N10} but as a function of the system size $N$ for a fixed noise strength $D=0.017$. Squares depict the simulation data corresponding to the full dynamics (\ref{eq:lang}). Lines are a guide to the eye.
\vspace{0.5em} 
}
\end{figure}

\begin{figure}
\includegraphics[width=8.5cm]{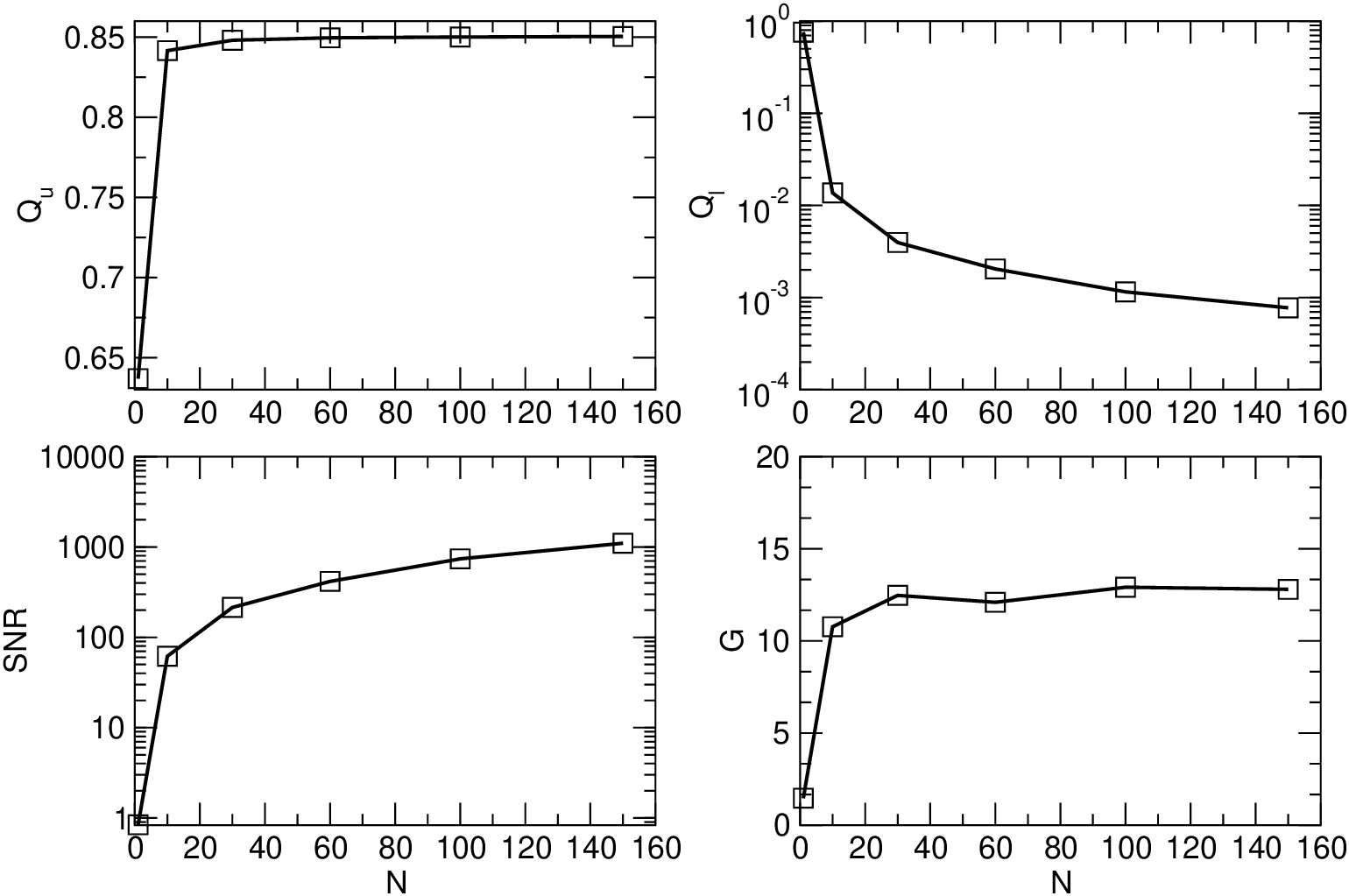}
\caption{
\label{fig:supra}
Suprathreshold dynamics. Same  as in Fig.~\ref{fig:sub} but for $D=0.2$.
}
\end{figure}

A numerical analysis of the deterministic version (i.e. without the noise terms $\eta_k$) of  Eqs.~(\ref{eq:third}) governing the third order approximation confirms that under a constant driving of $A=0.3$, there is a transition at about $0.02$ between a situation in which the system presents two attractors (subthreshold dynamics) and only one (suprathreshold dynamics). In fact, this analysis can also be carried out with the fluctuating cumulant theory presented in Sec.~\ref{sec:eff} with an arbitrary order of truncation, and thus, with an arbitrary order of accuracy. 

Let us illustrate the above discussion by considering two systems: one with a noise strength value $D=0.017$, which is just below the transition value $D_c(A\!=\!0.3)$; and one with $D=0.2$, well above that transition value. Figures \ref{fig:sub} and \ref{fig:supra} shows the behavior of the stochastic resonance quantifiers as a function of the system size $N$ for these two systems under the same rectangular driving with $A=0.3$ and $\Omega=0.01$. It can be seen that, for large enough $N$, the size of the fluctuations ($Q_l$) is always reduced as $N$ is increased, as expected. However, both systems display a very different behavior. In Fig.~\ref{fig:sub} the SR quantifiers $Q_u$, SNR and the gain are observed to decrease monotonically as $N$ is increased. This is what is expected for a system under subthreshold dynamics, because the driving force alone is unable to provoke jumps between the attractors and needs the presence of fluctuations. In the limit $N\rightarrow\infty$, the fluctuations, and thus the jumps, are completely suppressed. On the other hand, a very different behavior is observed in Fig.~\ref{fig:supra} for a noise strength value above the transition value corresponding to this driving amplitude. In this case, there is only one attractor, which is displaced to the positive or negative axis around the origin according to the instantaneous sign of the driving force. Here the presence of finite size fluctuations is only a nuisance to the driven oscillations of the collective variable. As a result, the spectral amplification and the SR gain grow with $N$ toward the finite values which correspond to the infinite system.  

Finally, let us notice that the phenomenon of {\em system size resonance} \cite{pikzai02}, i.e. the non-monotonic behavior of the SR quantifiers when plotted as a function of $N$, could only appear in a parameter region in which the system dynamics is subthreshold. Nevertheless, this phenomenon is not observed in Fig.~\ref{fig:sub}. This is due to the fact that the strength of the fluctuations is very small even for small values of the system size $N$, already below the optimal noise value in which the response is maximized. Any further increase of the system size reduces the fluctuations and, thus, the jumps between the attractors.

\section{Conclusions}
\label{sec:con}
Starting from the Langevin equations defining the model system, we have derived, using It\^o stochastic calculus, a hierarchy of exact stochastic differential equations for a set of fluctuating cumulant variables, defined by using the arithmetic mean over all oscillators. Due to the useful mathematical properties of It\^o stochastic calculus, the hierarchy contains noise terms with simple autocorrelation properties. Furthermore, the approach proposed in this paper for the fluctuating cumulants is not directly applicable with Stratonovich calculus. In the limit of an infinite number of oscillators, the whole hierarchy reduces to the one obtained by Desai and Zwanzig \cite{deszwa78} for the expected values of the cumulants. In contrast to the theory presented in Ref.\cite{deszwa78}, the fluctuating cumulant approach allows us to study a wide kind of collective dynamical properties like autocorrelation functions or the SNR, in addition to effects due to finite size fluctuations.  

Nevertheless, the noise terms that appear in the exact hierarchy for the fluctuating cumulant variables depend in a complicated way on the fluctuating cumulants and approximations have to be taken in order to obtain a closed set of stochastic differential equations. Here it is shown that this difficulty can be overcome in the asymptotic limit of a very large number of oscillators. However, one still has the inconvenience of dealing with a hierarchy with an infinite number of equations, and a truncation scheme is desirable. A Gaussian approximation was proposed by Pikovsky et al.~in Ref.~\cite{pikzai02}, and here it is presented as a second-order truncation scheme of the fluctuating cumulant hierarchy. In addition, an arbritrary-order truncation method is proposed, with explicit expressions given for the third order only. This third order approach turns out to provide the best quantitative agreement with the SR data. Finally, a rather simple approach based on a single variable and the use of an effective potential is proposed.
 
The spectral amplification of the collective variable as a function of the noise strength $D$ of systems with $N=10$ and $N=30$ bistable oscillators is found to be in good agreement with the predictions given by all the approximations, though small but appreciable systematic deviations are observed for the Gaussian approximation for a system with $N=30$ oscillators. Additionally, the effective potential theory and the Gaussian approximation do not account well for the SNR or the SR gain of the collective variable, though the data is within the same order of magnitude. A systematic underestimation of the fluctuations of the collective variable is done by the effective potential approach due to the mean-field like character of this simplified theory. The best quantitative agreement of the SNR and the SR gain is given by the third-order fluctuating cumulant theory, although a small systematic underestimation of the fluctuations is still observed with this third order theory due to the neglect of higher order fluctuating cumulants.

Furthermore, using any of the approximations presented, we are able to explain the very large gain values observed in Refs.~\cite{casgom06,cubcas07}. Specifically, it is shown that the driving amplitudes used are suprathreshold from the point of view of the effective dynamics in the range of noise strength values utilized in those works, i.e., there is only one attractor of the dynamics under the presence of the driving force, and this attractor oscillates following the driving force. Simulation results, showing several SR quantifiers as a function of the system size $N$, confirm this behavior. 

This situation resembles the effect of a high-frequency signal in an isolated bistable system. In the later case, the high-frequency signal can be removed from the description by means of an effective bistable potential with modified parameters, with the consequence that previously subthreshold driving amplitudes can become suprathreshold from the point of view of the effective potential \cite{cascub07}. In contrast, the effective dynamics induced by the high-frequency signal has been shown to provoke the opposite effect on an excitable system, being able to suppress the excitable character of the system \cite{cubbal06}. This suggests that much work is needed in order to extend the present analysis to finite sets of coupled excitable systems \cite{saihen06}. 

\begin{acknowledgments}

This research was supported by the Direcci\'on General de Ense\~nanza
Superior of Spain (Grant No. FIS2005-02884), the Junta de Andaluc\'{\i}a, and
the Juan de la Cierva program of the Ministerio de Ciencia y
Tecnolog\'{\i}a. 
The author thanks Jes\'us Casado-Pascual for helpful discussions.
\end{acknowledgments}

\appendix

\section{Derivation of Eq.~(\ref{eq:mukmukp})}
\label{sec:apen1}
In the notation commonly utilized within the framework of It\^o calculus, Eq.~(\ref{eq:mukmukp}) can be expressed as 
\begin{eqnarray}
\ud\Gamma_k(t)\,\ud\Gamma_l(t)&=&\frac{2D}{N}M_{k+l}(t)dt, \\
\ud\Gamma_k(t)\,\ud\Gamma_l(t^\prime)&=&0 \quad \mbox{for $t\ne t^\prime$},
\label{eq:apen1:gammak}
\end{eqnarray}
where $\Gamma_k(t)=\int_0^t\ud \tau\, \mu_k(\tau)$. To prove these equations we start from the definition (\ref{eq:muk:def}) to arrive at 
\begin{equation}
\ud\Gamma_k(t)=\frac{(2D)^{1/2}}{N}\sum_{i=1}^N[y_i(t)]^k\ud B_i(t),
\end{equation}
where $y_i(t)=x_i(t)-S(t)$. Thus
\begin{eqnarray}
\ud\Gamma_k(t)\,\ud\Gamma_l(t)&=&\frac{2D}{N^2}\sum_{i,j}[y_i(t)]^k[y_j(t)]^l\ud B_i(t) \ud B_j(t) \nonumber\\
&=& \frac{2D}{N^2}\sum_{i}[y_i(t)]^{k+l}dt \nonumber \\
&=&\frac{2D}{N}M_{k+l}(t)dt,
\end{eqnarray}
where we have used (\ref{eq:wienner}). Similarly, using the fact that the Wiener processes $B_i(t)$ have independent increments, i.e. $\ud B_j(t)\,\ud B_j(t^\prime)=0$ for $t\ne t^\prime$, and that the increments $\ud B_j(t^\prime)$ are independent of $y_i(t)$ at times $t^\prime\ge t$, Eq.~(\ref{eq:apen1:gammak}) is readily proven. 

In order to clarify the advantages of It\^o calculus in the context of this paper, let us now compute the autocorrelation of $\mu_k(t)$ by using Stratonovich calculus. In this case we are entitled to utilize the usual rules of differentiation of deterministic calculus. To that aim, the Novikov-Furutsu theorem \cite{fur63,nov65,han85} states that if $\xi(t)$ is a Gaussian white noise with zero mean and autocorrelation $\langle \xi(t)\xi(s)\rangle=2D\delta(t-s)$, then for any functional $g[\xi]$ we have
\begin{equation}
\langle \xi(t)g[\xi]\rangle=\int \ud s \langle \xi(t)\xi(s)\rangle  \langle \frac{\delta g[\xi]}{\delta \xi(s)}\rangle=2D\langle\frac{\delta g[\xi]}{\delta \xi(t)}\rangle,
\end{equation}
where $\delta g[\xi]/\delta \xi(t)$ denotes the functional derivative of $g[\xi]$. Thus, assuming $t\le t^\prime$,
\begin{eqnarray}
\langle \mu_k(t) \mu_{l}(t^\prime)\rangle&=&\frac{2D}{N^2}\sum_{i,j}\langle \frac{\delta[y_i(t)^k\xi_i(t) y_j(t^\prime)^l ]}{\delta\xi_j(t^\prime)}\rangle \nonumber \\
&=& \frac{2D}{N^2}\sum_{i,j}\Big\{ \langle \xi_i(t) y_i(t)^k \frac{\delta[y_j(t^\prime)^l ]}{\delta\xi_j(t^\prime)}\rangle+ \nonumber\\
& &\quad \langle \xi_i(t) y_j(t^\prime)^l \frac{\delta[y_i(t)^k ]}{\delta\xi_j(t^\prime)}\rangle+  \nonumber\\
& &\quad \langle y_i(t)^k y_j(t^\prime)^l \frac{\delta\xi_i(t)}{\delta\xi_j(t^\prime)}\rangle \Big\}.
\label{eq:muapen}
\end{eqnarray}
Using repeatedly the Novikov-Furutsu theorem and the fact that $\delta y_j(t)/\delta \xi_i(t)=(1/2)\delta_{ij}$, we arrive at Eq.~(\ref{eq:mukmukp}) plus the following two extra terms on the righthand side of Eq.~(\ref{eq:mukmukp})
\begin{equation}
D^2 l k\langle M_{k-1}(t)M_{l-1}(t^\prime)\rangle+\frac{2D^2l}{N^2}\sum_i\langle y_i(t)^k\frac{\delta y_i(t^\prime)^{l-1}}{\delta \xi_i(t)}\rangle.
\end{equation} 
These extra terms make the problem much more difficult to deal with.

\section{Gaussian noises in the limit of large number of oscillators}
\label{sec:apen2}
In this appendix we show that the process $\mu_k(t)$ tends to a Gaussian behavior as $N\rightarrow\infty$. First note that the third moment of $\mu_k(t)$,
\begin{equation}
\langle \mu_k(t_1) \mu_k(t_2)\mu_k(t_3) \rangle=0,
\end{equation}
and all odd moments of $\mu_k(t)$ vanish. If $\mu_k(t)$ were Gaussian, all cumulants higher than the second should be zero. This requires all odd moments of $\mu_k(t)$ to vanish but also a specific functional form of the even moments \cite{vankampen}. For instance, were $\mu_k(t)$ a Gaussian process, the fourth moment $\langle \mu_k(t_1) \mu_k(t_2)\mu_k(t_3) \mu_k(t_4)\rangle$ would be given by
\begin{eqnarray}
& &\langle \mu_k(t_1)\mu_k(t_2)\rangle \langle \mu_k(t_3)\mu_k(t_4)\rangle \nonumber\\
 & &+\, \langle \mu_k(t_1)\mu_k(t_3)\rangle \langle \mu_k(t_2)\mu_k(t_4)\rangle \nonumber\\
& &+\, \langle \mu_k(t_1)\mu_k(t_4)\rangle \langle \mu_k(t_2)\mu_k(t_3)\rangle.
\label{eq:muk4:1}
\end{eqnarray}
Instead, a simple calculation shows that for a finite $N$ the fourth moment is
\begin{eqnarray}
& &\langle \mu_k(t_1) \mu_k(t_2)\mu_k(t_3) \mu_k(t_4)\rangle = \left(\frac{2D}{N}\right)^2 \nonumber \\ 
& &\times \Big[ 
\langle M_{2k}(t_1)M_{2k}(t_2)\rangle \delta(t_1-t_4) \delta(t_2-t_3) \nonumber\\ 
& &\quad+\,\langle M_{2k}(t_1)M_{2k}(t_2)\rangle \delta(t_1-t_3) \delta(t_2-t_4) \nonumber\\
& &\quad+\,\langle M_{2k}(t_2)M_{2k}(t_3)\rangle \delta(t_1-t_2) \delta(t_3-t_4) \Big].
\label{eq:muk4:2}
\end{eqnarray}
Therefore, by considering (\ref{eq:mukmukp}), it is clear that (\ref{eq:muk4:1}) does not equal (\ref{eq:muk4:2}) unless
\begin{equation}
\langle M_{2k}(t_1) M_{2k}(t_2) \rangle=\langle M_{2k}(t_1) \rangle \langle M_{2k}(t_2) \rangle.
\label{eq:m2gauss}
\end{equation}
This identity is asymptotically correct in the limit $N\rightarrow \infty$, because then all fluctuating moments become deterministic. For a large enough $N$, Eq.~(\ref{eq:m2gauss}) will hold as a good approximation, showing a Gaussian behavior in the lowest order of a $N^{-1}$ expansion. Clearly, similar considerations apply to other even moments of $\mu_k(t)$. 


\end{document}